\documentclass[aps,prd,onecolumn,amssymb]{revtex4}
\usepackage{graphicx,bm}
\usepackage[english]{babel}
\usepackage{amsmath}
\usepackage{amsfonts}
\begin{document}

\newcommand{\be}{\begin{equation}}
\newcommand{\ee}{\end{equation}}
\newcommand{\bea}{\begin{eqnarray}}
\newcommand{\eea}{\end{eqnarray}}
\newcommand{\ben}{\begin{enumerate}}
\newcommand{\een}{\end{enumerate}}
\newcommand{\nn}{\nonumber \\}
\newcommand{\e}{\mathrm{e}}

\title{Brane cosmology from observational surveys and its comparison \\ with standard FRW cosmology}
\author{A.~V. Astashenok$^{a,}$\footnote{E-mail: artyom.art@gmail.com},\ E. Elizalde$^{b,}$\footnote{E-mail: elizalde@ieec.uab.es, elizalde@math.mit.edu},\ J. de Haro$^{c,}$\footnote{E-mail: jaime.haro@upc.edu}, \
S.~D. Odintsov$^{b,d,e}$\footnote{E-mail: odintsov@ieec.uab.es},\ A.~V. Yurov$^{a,}$\footnote{E-mail: yurov@freemail.ru}}
\affiliation{$^{a}$Baltic Federal University of I. Kant, Department of Theoretical Physics,
 236041, 14 Nevsky St., Kaliningrad, Russia \\
 $^{b}$Consejo Superior de Investigaciones Cient\'{\i}ficas, ICE/CSIC and IEEC, \\
 Campus UAB, Facultat de Ci\`encies, Torre C5-Par-2a pl, 08193 Bellaterra (Barcelona) Spain\\
$^c$Departament de Matem\`atica Aplicada I, Universitat Polit\`ecnica de Catalunya,\\
Diagonal 647, 08028 Barcelona, Spain\\
$^d$Instituci\'{o} Catalana de Recerca i Estudis Avan\c{c}ats (ICREA)\\
$^e$Eurasian National University, Astana, Kazakhstan and Tomsk State Pedagogical University, Tomsk, Russia}
\begin{abstract}
Several  dark energy models on the brane are investigated. They are compared with corresponding theories in the frame of 4d Friedmann-Robertson-Walker cosmology. To constrain the parameters of the models considered, recent observational data, including SNIa apparent magnitude measurements, baryon acoustic oscillation results, Hubble parameter evolution data and matter density perturbations are used. Explicit formulas of the so-called {\it state-finder} parameters
in teleparallel theories are obtained that could be useful to test these models and to establish a link between Loop Quantum Cosmology and Brane Cosmology. It is concluded that a joint analysis as the one developed here allows to estimate, in a very convenient way, possible deviation of the real universe cosmology from the standard Friedmann-Robertson-Walker one.
\end{abstract}


\maketitle

\section{Introduction}

A number of difficult problems in cosmology have been put forward by the discovery of the accelerated expansion of the universe
\cite{Riess,Perlmutter}.
This cosmic acceleration  can be explained via the introduction of  dark energy (for a recent review, see \cite{Dark-6}).
It follows from recent observational results that dark energy currently accounts for about 73\% of the total mass/energy of the universe \cite{Kowalski}. It appears to have rather strange properties, as a negative pressure and/or negative entropy, the fact that it is undetectable in the early universe,
and so on. It is still not excluded, however, that General Relativity (GR) and the ensuing
 vacuum fluctuations (as those leading, e.g., to the Casimir effect)
could lead to an explanation of the issue (see e.g. \cite{vari1,Book}).
One should also stress the following important connection: with the help
of an ideal fluid, GR can actually be rewritten, in an equivalent way,
as some modified gravity (for a recent review, see \cite{Nojiri}).

For dark energy with density $\rho_\mathrm{D}$ and pressure $p_\mathrm{D}$, the equation of state (EoS) parameter $w_\mathrm{D}$,
\be
w_\mathrm{D}=p_\mathrm{D}/\rho_\mathrm{D}<0\, ,
\ee
is known to be negative. Moreover, astrophysical observations clearly favor, up to now, the standard $\Lambda$CDM cosmology. Dark energy as just a cosmological constant ($w_{D}=-1$) is the simplest and maybe best preferred model from the theoretical viewpoint, too. In this model over 70\% of the current energy budget is dark energy (Einstein's cosmological constant $\Lambda$), in perfect accordance with the data coming from observations, as reported above.

Presently several independent observational procedures provide strong evidence in favor of the $\Lambda$CDM model, in particular SNIa apparent magnitude measurements as a function of the redshift, cosmic microwave background (CMB) anisotropies, baryon acoustic oscillation (BAO) peak length scale measurements, Hubble parameter determinations etc., but the error bars associated with all these classes of data are still too large to allow
for a significant observational discrimination between the $\Lambda$CDM model and other existing, theoretically well grounded alternatives to it.

When $w<-1$ (phantom dark energy) \cite{Caldwell}, we face up the most interesting and less understood theoretical possibility. A simultaneous violation of all four energy conditions occurs in this case and the involved field is unstable, although it could perhaps be made stable in classical cosmology \cite{Carrol}. For a universe filled with phantom energy there are many possible new scenarios for the end of such universe, among which the most typical cases are those of a Big Rip singularity
\cite{Caldwell, Frampton,S,BR,Nojiri} and of a sudden future singularity \cite{Barrow}. However, a final evolution without singularities is also possible: if the parameter $w$ asymptotically tends to $-1$, and the energy density increases with time or remains constant,  no finite-time future singularity will be ever formed
\cite{Sahni0,Frampton-2,Frampton-3,Astashenok,LR}, also, if the universe starts to decelerate in the far future. In any case, if the energy density grows up to some threshold value, the disintegration of any bound structure will eventually occur, in a way quite similar to the case of the Big Rip singularity, but this may only happen faraway in the future evolution.
The  dark energy pressure is expressed as a function of the density, as
\be
\label{EoS-0}
p=g(\rho).
\ee

In this paper a number of dark energy models on the brane will be considered (for a general introduction to brane-world cosmology, see~\cite{roy}). The theoretical predictions of these models will be compared with the various types of existing independent data observations, including the luminosity distance modulus vs redshift for SNe Ia, the data accumulated on the evolution of the Hubble parameter $H(z)$, the latest baryon acoustic oscillation (BAO) results, and matter density perturbation data. In Sect.~II a brief overview of the EoS fluid formalism is presented. A comparison of dark energy in Friedmann-Robertson-Walker (FRW) cosmology and on the brane is carried out. The main constraints coming from the observational survey data will be analyzed in Sect.~III. In the following two sections, Sects.~IV and V, we will study the simplest $\Lambda$CDM model on the brane  and will show that a careful joint analysis of the various observational data allows to estimate, in a clear fashion, any possible deviation of our cosmological model from the standard FRW cosmology.
Due to the increasing interest in teleparallel theories ($F(T)$ models), we deal with them in Sect.~VI where we find explicit formulas for the so-called {\it state-finder} parameters in such theories, which could be useful to test the models proposed there, in particular to test Loop Quantum Cosmology and Brane Cosmology, and specific relations between the two. Finally, Sect.~VII is devoted to conclusions.

\section{Comparison of dark energy in FRW- and brane-cosmology}

We start with a brief description of dark energy models in the frame of the FRW cosmology. The cosmological equations corresponding to a spatially flat universe, endowed with a metric
\be
ds^{2}=-dt^{2}+a^{2}(t)(dx^{2}+dy^{2}+dz^{2}),
\ee
are the following
\be
\label{Fried1}
\left(\frac{\dot a}{a}\right)^2 =  \frac{\rho}{3}\, , \quad
\dot{\rho} =  -3\left(\frac{\dot a}{a}\right)(\rho + p)\, .
\ee
with $\rho$ and $p$, respectively, the total energy-density and pressure,
while $a$ is the scale factor, the dot means time derivative,
and natural system units are being used, with $8\pi G=c=1$.

For dark energy, the EoS can be rewritten, for convenience, in the form
\be \label{EoS}
p_{D}=-\rho_{D}-f(\rho_{D}),
\ee
being $f(\rho_{D})$ a function of the energy-density. We observe that $f(\rho_{D})>0$ corresponds to $w<-1$, and $f(\rho_D)<0$ to $w>-1$.
The future evolution of the universe depends on the EoS for dark energy chosen. Let us here describe two main cases. \medskip

\noindent {\sf (i) Evolution without future singularities.} This case includes a so-called ``Little Rip'' (these models are described in detail in \cite{Frampton-2,Frampton-3}). The dark energy density grows with time so slowly that a Big Rip cannot occur in finite time. For the realization of this scenario, one needs the asymptotic behavior of the function to be $g(\rho_{D})\sim \rho_{D}^{\beta}$, as $\beta \leq 1/2$. Eventually, a dissolution of all bound structures will also take place in the future.

One should note that, for some specific equations of state with branch points, a de(phantomization) process can occur \cite{Wei-0}. Therefore, after the acceleration phase, a slowdown of the future universe might be possible. In other words, the universe may be decelerating in the future.

Interesting alternatives to the $\Lambda$CDM model are models in which the dark energy density tends asymptotically to a constant value (an ``effective cosmological constant'' \cite{Astashenok}). One should remark that if the value of this ``effective cosmological constant'' is sufficiently large (for example, if $\Lambda_{eff}\sim 1$ in Planck units) a possibility of disappearance of bound structures due to the enormous acceleration of the universe still remains. \medskip

\noindent {\sf (ii) Evolution with finite-time singularities.} If $g(\rho_{D})\sim \rho_{D}^{\beta}$, with $\beta>1/2$, the dark energy density grows so rapidly that the universe ends its existence in a singularity of Big Rip type or in a type III singularity, according to the classification in \cite{Nojiri-2}. The key difference between these singularities is that the energy density in the second case grows so rapidly with time that the scale factor does never reach the infinite value. These scenarios can be realized only in the case of having a phantom energy. Another interesting case occurs if $f(\rho_{D})\rightarrow\pm \infty$ at $\rho_{D}=\rho_{Df}$, i.e., the dark energy pressure becomes infinite at finite energy density. The second derivative of the scale factor diverges, while the first derivative remains finite.

As an alternative to the FRW cosmology let us consider the simplest brane model in which spacetime is homogeneous and isotropic along three spatial dimensions, being our 4-dimensional universe an infinitesimally thin wall, with constant spatial curvature, embedded in a 5-dimensional spacetime \cite{Sahni,Langlois}. In the Gaussian normal coordinate system, for the brane which is located at $y=0$, one has
\be
ds^{2}=-n^{2}dt^{2}+a^{2}(t,y)\gamma_{ij}dx^{i}dx^{j}+\epsilon dy^{2},
\ee
where $\gamma_{ij}$ is the maximally 3-dimensional metric, and $\epsilon=1$, if the extra dimension is space-like, while $\epsilon=-1$, if it is time-like.
Let $t$ be the proper time on the brane ($y=0$), then $n(t,0)=1$. Therefore, one gets the FRW metric on the brane
\be
ds^{2}_{|y=0}=-dt^{2}+a^{2}(t,0)\gamma_{ij}dx^{i}dx^{j}.
\ee
The 5-dimensional Einstein equations have the form
\be
R_{AB}-\frac{1}{2}g_{AB}R=\chi^{2}T_{AB}+\Lambda g_{AB},
\ee
where $\Lambda$ is the bulk cosmological constant. The next step is to write the total energy momentum tensor $T_{AB}$ on the brane as
\be
T^{A}_{B}=S^{A}_{B}\delta(y),
\ee
with $S^{A}_{B}=\mbox{diag}(-\rho_{b},p_{b},p_{b},p_{b},0)$, where $\rho_{b}$ and $p_{b}$ are the total brane energy density and pressure, respectively.

One can now calculate the components of the 5-dimensional Einstein tensor which solves Einstein's equations. One of the crucial issues here is to use appropriate junction conditions near $y=0$. These reduce to the following two relations:
\be
\frac{dn}{ndy}_{|y=0+}=\frac{\chi^{2}}{3}\rho_{b}+\frac{\chi^{2}}{2}p_{b},\qquad \frac{da}{ady}_{|y=0+}=-\frac{\chi^{2}}{6}\rho_{b}.
\ee
After some calculations, one  obtains the following result
\be
H^{2}=\epsilon\chi^{4}\frac{\rho_{b}^{2}}{36}+\frac{\Lambda}{6}-\frac{k}{a^{2}}+\frac{C}{a^{4}}.
\ee
This expression is valid on the brane  only. Here $H=\dot{a}(t,0)/a(t,0)$ and $C$ is an
arbitrary integration constant. The energy conservation equation is correct, too,
\be
\dot{\rho_{b}}+3\frac{\dot{a}}{a}(\rho_{b}+p_{b})=0.
\ee
Now, let $\rho_{b}=\rho+\lambda$, where $\lambda$ is the brane tension. For a fine-tuned brane with $\Lambda=\epsilon\lambda^{2}\chi^{4}/6$, we have the equation (for $k=0$)
\be \label{BREQ-1}
\frac{\dot{a}^{2}}{a^{2}}=\frac{\epsilon\lambda\chi^{4}}{6}\frac{\rho}{3}
\left(1+\frac{\rho}{2\lambda}\right)+\frac{C}{a^{4}}.
\ee
In what follows we will consider a single brane model which mimics GR at present but differs from it at late times.
We set $8\pi G=\epsilon\sigma\chi^4/6$. As can be checked, two cases arise: $\epsilon=1$ and $\lambda>0$, and $\epsilon=-1$ and $\lambda<0$. For simplicity, we also set $C=0$ (the term with $C$ is usually called ``dark radiation''). In fact, allowing $C\neq 0$ does not lead
to additional solutions on a radically new basis, in the framework of our approach. Eq.~(\ref{BREQ-1}) can be simplified to
\be \label{BREQ}
\frac{\dot{a}^{2}}{a^{2}}=\frac{\rho}{3}\left(1+\frac{\rho}{2\lambda}\right).
\ee
 Eq.~(\ref{BREQ}), for $\rho<<|\lambda|$, differs insignificantly from the FRW equation.  The brane model with a positive tension has been discussed in \cite{Liddle,Sami,Sami-2} in the context of  the unification of early- and late-time acceleration eras.
The braneworld model with a negative tension and a time-like extra dimension can be regarded as being dual to the Randall-Sundrum model \cite{Sahni-2,Randall,Copeland} which, as we will show in Sect.~VI A, is in fact equivalent to the effective formulation of loop quantum cosmology.
Note also that, for this model, the Big Bang and Rip singularities are
absent (see Sect.~VI A). And this fact does not depend upon whether or not matter violates the energy conditions \cite{Asht}. This same scenario has also been used to construct cyclic models for the universe \cite{Cyclic}.

One can assume that in our epoch $\rho/2\lambda<<1$ and thus there is no significant difference between the brane model and FRW cosmology. But the universe evolution in the future, for brane cosmology, can in fact differ from such convenient cosmology, due to the non-linear dependence of the expansion rate on the energy density.

The EoS formalism for dark energy models on the brane was considered in \cite{Brane}. Here we briefly describe this approach. One gets the following link between time and dark energy density, assuming that $\rho_{D}>>\rho_{m}$:
\begin{equation}\label{brtrho}
t(\rho_{D})-t_{0}=\frac{1}{\sqrt{3}}\int_{\rho_{D0}}^{\rho_{D}}\frac{d\rho}{\rho^{1/2}\left(1+\frac{\rho}{2\lambda}\right)^{1/2}f(\rho)}.
\end{equation}
For the present time, $t_{0}$, we can set $t_{0}=0$. For the scale factor as a function of the dark energy density, we have the same relation as in the FRW cosmology, namely
\begin{equation}\label{arho}
a=a_{0}\exp\left(\frac{1}{3}\int_{\rho_{D0}}^{\rho_{D}}\frac{d\rho}{g(\rho)}\right).
\end{equation}

In the case of a positive tension, the following possibilities can be realized:
\ben
\item  If the integral (\ref{brtrho}) converges while (\ref{arho}) diverges, we have a Big Rip. It is interesting to note that the Big Rip on a brane considered in \cite{Yurov} occurs faster than in the FRW cosmology.

For the simplest EoS with constant state parameter $w_{0}=-1-\alpha^{2}$, the function $g(\rho_{D})=\alpha^{2}\rho_{D}$. If $\rho_{D}>>\lambda$, then the dark energy density grows with time substantially faster than in  ordinary cosmology ($\lambda\rightarrow\infty$).
\item  If the integrals (\ref{brtrho}) and (\ref{arho}) diverge when $\rho_{D}\rightarrow\infty$, then a  Little Rip occurs. The acceleration of the universe increases with time definitely faster than in the FRW universe, owing to the brane  tension (see the corresponding time equation for the case (1)).
\item  Asymptotic de Sitter expansion is realized if $g\rightarrow 0$ for $\rho_{D}\rightarrow\rho_{Df}$, and the integral (\ref{brtrho}) diverges.
\item  There is a type III singularity if both integrals converge when $\rho_{D}\rightarrow\infty$.
\item  If $g(\rho_{D})\rightarrow\infty$ for $\rho_{D}\rightarrow\rho_{Df}$, the universe ends its existence in a sudden future singularity.
\een
The case of negative tension allows for the following interesting possibilities:
\ben
\item  Asymptotic de Sitter expansion, if $g(\rho_{D})\rightarrow0$ for $\rho_{D}\rightarrow\rho_{Df}$.
\item An asymptotic breakdown (i.e. the rate of expansion of universe tends to 0) will occur  if $g(\rho_{D})\rightarrow 0$ for $\rho_{D}\rightarrow2\lambda$.
\item A sudden future singularity, if $f(\rho_{D})\rightarrow\infty$ when $\rho_{D}\rightarrow\rho_{Df}$.
\een
One should note that dark energy with an EoS such that $f(\rho)\sim\rho^{\gamma}$, with $\gamma\leq2$, leads to a Big Rip on the brane while, in the case of the conventional FRW universe, such dark energy leads to a Little Rip only.

\section{Observational data}

The parameters of the cosmological models can be determined from a strict comparison of their predictions with accurate observational data. We here consider the data coming from SNe observations, the evolution of the Hubble parameter, baryon acoustic oscillation, and the evolution of matter perturbations.

\subsection{SNe observations.} The modulus $\mu$ vs redshift $z=a_{0}/a-1$ relation corresponding to type Ia supernovae from the Supernova Cosmology Project \cite{Amanullah},\cite{Union2} is, as well known,
\be
\mu(z)=\mu_{0}+5\lg D_{L}(z).
\ee
The relation for the luminosity distance $D_{L}(z)$ as a function of the redshift in the FRW cosmology (FC) is
\be \label{DLFC}
D^{FC}_{L}=\frac{c}{H_{0}}(1+z)\int_{0}^{z}
h^{-1}(z)dz,\quad h(z)=\left[\Omega_{m0}(1+z)^{3}+\Omega_{D0}F(z)\right]^{1/2}
\ee
Here, $\Omega_{m0}$ is the total fraction of matter density, $\Omega_{D0}$ the fraction of dark energy energy density, and $H_{0}$ the current Hubble parameter. The constant value $\mu_{0}$ depends on the chosen Hubble parameter:
$$
\mu_{0}=42.384-5\log h,\quad h=H_{0}/100 \mbox{km/s/Mpc}
$$
The function $F(z)=\rho_{D}(z)/\rho_{D0}$ can be determined from the continuity equation
\begin{equation}
\dot{\rho}_{D}-3\frac{\dot{a}}{a}g(\rho_{D})=0,
\end{equation}
which can be rewritten as
\begin{equation}
\int_{\rho_{D0}}^{\rho_{D}(z)}\frac{dy}{g(y)}=-3\ln(1+z).
\end{equation}
For simplicity, we neglect the contribution of radiation.








For cosmology on the brane (BC), Eq.~(\ref{DLFC}) can be rewritten as
\be \label{DLBC}
D^{BC}_{L}=\frac{c}{H_{0}}(1+z)\int_{0}^{z}
h^{-1}(z)\left[1+\delta h^{2}(z)\right]^{-1/2}(1+\delta)^{1/2}d z\,
\ee
where for convenience the parameter $\delta=\rho_{0}/2\lambda$ has been introduced. For the analysis of the SNe data one needs to calculate the parameter $\chi^{2}$, which is defined by
\begin{equation}
\chi^{2}_{SN}=\sum_{i}\frac{\left[\mu_{obs}(z_{i})-
\mu_{th}(z_{i})\right]^{2}}{\sigma^{2}_{i}},
\end{equation}
where $\sigma_{i}$ is the corresponding $1\sigma$ error. The parameter $\mu_{0}$ is independent of the data points and, therefore, one has to perform a uniform marginalization over $\mu_{0}$. Minimization with respect to $\mu_{0}$ can be done by simply expanding the $\chi^{2}_{SN}$ with respect to $\mu_{0}$,
\begin{equation}\label{chi}
\chi^{2}_{SN}=A-2\mu_{0}B+\mu_{0}^{2}C,
\end{equation}
where
$$
A=\sum_{i}\frac{\left[\mu_{obs}(z_{i})-\mu_{th}(z_{i};\mu_{0}=0)\right]^{2}}{\sigma^{2}_{i}},
$$
$$
B=\sum_{i}\frac{\mu_{obs}(z_{i})-\mu_{th}(z_{i})}{\sigma^{2}_{i}},\quad C=\sum_{i}\frac{1}{\sigma^{2}_{i}}.
$$
The expression (\ref{chi}) has a minimum for $\mu_{0}=B/C$ at
$$
\bar{\chi}_{SN}^{2}=A-B^{2}/C.
$$
One can minimize $\bar{\chi}_{SN}^{2}$ instead of ${\chi}_{SN}^{2}$. Following \cite{Nesseris}, one determines the 68.3\% confidence level (CL) by $\Delta\chi^{2}=\chi^{2}-\chi^{2}_{min}<1.0$ for the one-parameter or $2.3$ for the two-parameter model. Similarly, the 95.4\% confidence level is determined by $\Delta\chi^{2}=\chi^{2}-\chi^{2}_{min}<4.0$ or $6.17$ for the one- and two-parameter models, respectively.

\subsection{Hubble parameter.} The evolution of the Hubble parameter with time in the past is now well observed. The Hubble parameter depends on the differential age of the universe as a function of the redshift, in the form
$$
dt=-\frac{1}{H}\frac{dz}{1+z}.
$$
Therefore, a determination of $dz/dt$ directly measures $H(z)$. This is made possible through data we have on the absolute age of passively evolving galaxies, determined from fitting stellar population models. We use the 11 datapoints for H(z) from \cite{Stern} for constraining the model parameters. These data are listed in Table I. The theoretical dependence of the Hubble parameter in the brane model is
\begin{equation}
H(z)=H_{0}h(z)(1+\delta h^{2}(z))^{1/2}(1+\delta)^{-1/2}.
\end{equation}

\begin{table}
\label{Table1}
\begin{centering}
\begin{tabular}{|c|c|c|c|}
  \hline
  $z$ & $H_{obs}(z)$ & $\sigma_{H}$  \\
      & km s$^{-1}$ Mpc$^{-1}$        &  km s$^{-1}$ Mpc$^{-1}$     \\
  \hline
  0.090 & 69 & 12 \\
  0.170 & 83 & 8  \\
  0.270 & 77 & 14 \\
  0.400 & 95 & 17 \\
  0.480 & 97 & 62 \\
  0.880 & 90 & 40 \\
  0.900 & 117 & 23 \\
  1.300 & 168 & 17 \\
  1.430 & 177 & 18 \\
  1.530 & 140 & 14 \\
  1.750 & 202 & 40 \\
  \hline
\end{tabular}
\caption{Hubble parameter versus redshift data from \cite{Stern}.}
\end{centering}
\end{table}

The parameter $\chi^{2}_{H}$ is here
\begin{equation}
\chi^{2}_{H}=\sum_{i}\frac{\left[ H_{obs}(z_{i})-H_{th}(z_{i})\right]^{2}}{\sigma^{2}_{i}}.
\end{equation}
We need to perform a uniform marginalization over the parameter $H_{0}$. Again, we can expand
$$
\chi^{2}_{H}=A_{1}-2B_{1}H_{0}+H_{0}^{2}C_{1},
$$
$$
A_{1}=\sum_{i}\frac{H_{obs}(z_{i})^{2}}{\sigma^{2}_{i}},\quad B_{1}=\sum_{i}\frac{E(z_{i})H_{obs}(z_{i})}{\sigma^{2}_{i}},\quad C_{1}=\sum_{i}\frac{1}{\sigma^{2}_{i}}
$$
The parameter $\chi^{2}_{H}$ has a minimum at the point $H_{0}^{2}=B_{1}/C_{1}$,
$$
\bar{\chi}_{H}^{2}=A_{1}-B_{1}^{2}/C_{1}.
$$
As in the case of the SNe data, we could minimize $\bar{\chi}_{H}^{2}$ instead of ${\chi}_{H}^{2}$.

\subsection{BAO data.} To constrain cosmological parameters using BAO data we follow the procedure described in \cite{Blake}. We use the measurements of the acoustic parameter $A(z)$ from \cite{Blake}, where the theoretically predicted $A_{th}(z)$ is given by the relation
\begin{equation}\label{Ath}
A_{th}(z)=\frac{D_{V}(z)H_{0}\sqrt{\Omega_{m0}}}{z},
\end{equation}
where $D_{V}(z)$ is a distance parameter defined as
\begin{equation} \label{DV}
D_{V}(z)=\left[(1+z)^{2}d_{A}^{2}(z)\frac{cz}{H(z)}\right]^{1/3}.
\end{equation}
Here, $d_{A}(z)$ is the angular diameter distance
\begin{equation} \label{dA}
d_{A}(z)=\frac{y(z)}{H_{0}(1+z)},\quad y(z)=\int_{0}^{z}\frac{dz}{E(z)},\quad E(z)=H(z)/H_{0}.
\end{equation}

Using Eqs.~(\ref{Ath})-(\ref{dA}), we have
\begin{equation}
A_{th}(z)=\sqrt{\Omega_{m0}}\ \frac{y^{2}(z)}{z^{2}E(z)},
\end{equation}
and using the WiggleZ $A_{obs}(z)$ data from Table 3 of \cite{Blake}, we compute $\chi^{2}_{A}$ to be
\begin{equation}
\chi^{2}_{A}=\Delta \textbf{A}^{T}(C_{A})^{-1}\Delta\textbf{A}.
\end{equation}
Here, $\Delta\textbf{A}$ is a vector consisting of differences, $\Delta A_{i}=A_{th}(z_{i})-A_{obs}(z_{i})$ and $C^{-1}_{A}$ is the
inverse of the $3 \times 3$ covariance matrix given in Table 3 of \cite{Blake}.

\subsection{Matter density perturbations.} As was shown in \cite{Christopherson} one can neglect the density perturbations of dark energy. In this case the dark matter perturbations effectively decouple from DE perturbations. The equation that determines the evolution of the density contrast $\delta$ in a flat background filled by matter with density $\rho_{m}$ is
\be\label{Perturb}
\ddot{\delta}_{m}+2H\dot{\delta}_{m}=\frac{1}{2}\rho_{m}\delta_{m}.
\ee
It is convenient to introduce the  growth rate function of the perturbations $f=d\ln\delta_{m}/d\ln a$. Using the FRW equations, one gets the following equation for $f$
\be
\frac{df}{d\ln a}+f^{2}+\left(\frac{\dot{H}}{H^{2}}+2\right)f-\frac{3}{2}\Omega_{m}(1+\rho/2\lambda)^{-1}=0,
\ee
where $\Omega_{m}$ is the matter fraction of the total energy-density $\Omega_{m}=\Omega_{m0}(1+z)^{3}/[\Omega_{m0}(1+z)^{3}+\Omega_{D0}F(z)]$. Finally, using the relation
$$
\frac{d}{d\ln a}=-(1+z)\frac{d}{dz}
$$
and taking into account that
$$
\frac{\ddot{a}}{a}-\frac{\dot{a}^{2}}{a^{2}}=-\frac{\rho+p}{2}(1+\rho/\lambda),
$$
we get
\be \label{Perturb-2}
-(1+z)\frac{df}{dz}+f^{2}+\left(2-\frac{3\Delta}{2}\Omega_{m}+\frac{3\Delta}{2}\frac{g(\rho_{D})}{\rho}\right)f-
\frac{3}{2}\Omega_{m}(1+\rho/2\lambda)^{-1}=0.
\ee
where $\Omega_{D}=\rho_{D}/\rho$ and we have introduced the parameter $\Delta=(1+\rho/\lambda)(1+\rho/2\lambda)^{-1}$.

For a dark fluid with given EoS, one can find the DE density as a function of the redshift $z$. Then, Eq.~(\ref{Perturb-2}) can be solved numerically. The observational data for the  growth factor $f_{obs}$ at various redshifts are given in Table II.

\begin{table}
\label{Table2}
\begin{centering}
\begin{tabular}{|c|c|c|}
  \hline
  $z$ & $f_{obs}$ & Ref. \\
  \hline
  0.15 & $0.51\pm0.11$ & \cite{Hawkins}, \cite{Verde} \\
  0.32 & $0.654\pm0.18$ & \cite{Reyes} \\
  0.35 & $0.70\pm0.18$ & \cite{Tegmark} \\
  0.55 & $0.75\pm0.18$ & \cite{Ross} \\
  0.77 & $0.91\pm0.36$ & \cite{Guzzo} \\
  1.4 & $0.90\pm0.24$ & \cite{Angela} \\
  3.0 & $1.46\pm0.29$ & \cite{Mcdonald}\\
  \hline
\end{tabular}
\caption{Available data for the growth factor $f_{obs}$ at various redshifts from the change of the power spectrum Ly-$\alpha$ forest data in SDSS.}
\end{centering}
\end{table}

\section{$\Lambda$CDM model on the brane}

First, we consider the very simple cosmological model on the brane with vacuum energy $\rho_{D}=\Lambda=\mbox{const}$. This model coincides  in the future with the FRLW cosmology with a redefined cosmological constant. The asymptotic behavior of the scale factor is
\begin{equation}
a(t)\sim a_{0}\exp\left[({\Lambda_{eff}/3})^{1/2}t\right],\quad t\rightarrow\infty, \quad \Lambda_{eff}=\Lambda(1+\Lambda/2\lambda).
\end{equation}

\begin{table}
\label{Table1}
\begin{centering}
\begin{tabular}{|c|c|c|c|c|c|c|c|}
\hline
Data sets & SNe & H & BAO & SNe+H+BAO & SNe+H+BAO+F & $\chi^{2}_{min}$ \\
\hline
  $\delta=0$ & $0.722^{-0.019,-0.039}_{+0.020,+0.039}$ & $0.735^{-0.089,-0.213}_{+0.066,+0.117}$ & $0.699^{-0.028,-0.059}_{+0.026,+0.049}$ & $0.712^{-0.011,-0.027}_{+0.016,+0.031}$ & $0.712^{-0.009,-0.025}_{+0.018,+0.033}$ & 562.39  \\
  \hline
  $\delta=0.05$ & $0.744^{-0.016,-0.036}_{+0.016,+0.034}$ & $0.787^{-0.058,-0.130}_{+0.047,+0.086}$ & $0.686^{-0.031,-0.065}_{+0.028,+0.054}$ & $0.729^{-0.012,-0.028}_{+0.013,+0.028}$ & $0.729^{-0.012,-0.027}_{+0.013,+0.027}$ & 565.91  \\
  \hline
  $\delta=0.10$ & $0.759^{-0.013,-0.030}_{+0.014,+0.032}$ & $0.813^{-0.046,-0.102}_{+0.039,+0.072}$ & $0.673^{-0.028,-0.072}_{+0.031,+0.059}$ & $0.743^{-0.014,-0.028}_{+0.011,+0.025}$  & $0.743^{-0.014,-0.028}_{+0.010,+0.023}$  & 571.38 \\ \hline
\end{tabular}
\caption{Best fitting values for $\Omega_{\Lambda}$ within 1$\sigma$ and 2$\sigma$ errors for various $\delta$ from observational data analysis for the $\Lambda$CDM model on the brane. In the last column the minimal value of $\chi^{2}_{min}=
\bar{\chi}^{2}_{SN}+\bar{\chi}^{2}_{H}+\chi^{2}_{A}+\chi^{2}_{f}$ is given.}
\end{centering}
\end{table}

One can consider the $\Lambda$CDM model on the brane as a one-parametric model, at fixed values of $\delta$. The results of the calculations corresponding to this case are given in Table III (we have also included in our considerations the case $\delta=0$, i.e. FRW cosmology, for comparison).
The BAO data favor smaller values of $\Omega_{D0}$ than the $H(z)$ and SNe data. The optimal value of $\Omega_{\Lambda}$ (that is, $\Omega_{D}$ when $\rho_D=\Lambda$) is closer to the one coming from the SNe data analysis only. One easily sees that the addition of the observational data for the matter density perturbations does not change the best-fit value of $\Omega_\Lambda$ from the SNe+H+BAO analysis. One can also conclude that the best consistent description of all observational data is realized in the frame of the FRW cosmology ($\delta=0$ or $\lambda\rightarrow\infty$). The minimal value of the total $\chi^{2}$ is $562.39$. As $\delta$ grows the corresponding $\chi^{2}$ increases.

For $\delta>0$ we have the following picture. The analysis of the data sets does not yield separately a significant constraint on the maximal value of $\delta$. The parameter $\chi^{2}$ for SNe, BAO and matter density perturbation data grows very slowly with increasing $\delta$. For instance, for $\delta=0$, $\bar{\chi}^{2}_{SN, min}=553.18$, while for $\delta=0.1$ the minimal value of $\chi^{2}_{SN}$ is
$553.34$. The data on the evolution of the Hubble parameter are more sensitive to increasing $\delta$: for $\delta=0.1$ we found that
$\bar{\chi}^{2}_{H,min}=8.12$, in comparison with $\chi^{2}_{H,min}=7.62$ in the FRW model.
But one can see that, at $\delta\approx 0.05$, the $1\sigma$ intervals of the possible values
of $\Omega_\Lambda$ for SNe and BAO data do not intersect. The $2\sigma$ intervals for $\Omega_\Lambda$ from these data sets do not have common points for $\delta\approx 0.10$. Therefore, one can estimate the maximal value of $\delta$ from the joint analysis of all observational data sets.

For doing this, we consider the $\Lambda$CDM model on the brane as a two-parametric one, with free parameters $\delta$ and $\Omega_\Lambda$. One can see that, although the areas corresponding to the $1\sigma$ and $2\sigma$ contours from the SNe, BAO and $H(z)$ data analysis are sufficiently large
(see Fig.~1), these contours intersect in a quite narrow region of the parameter space. Joint data analysis allows us to define the $1\sigma$ and $2\sigma$ contours in the $\Omega_\Lambda$ - $\delta$ parameter space (Fig. 2). Therefore, we can estimate the upper limit of the parameter $\delta$ at which the $\Lambda$CDM model is relevant to the observational data.

\begin{figure}
\includegraphics[scale=1]{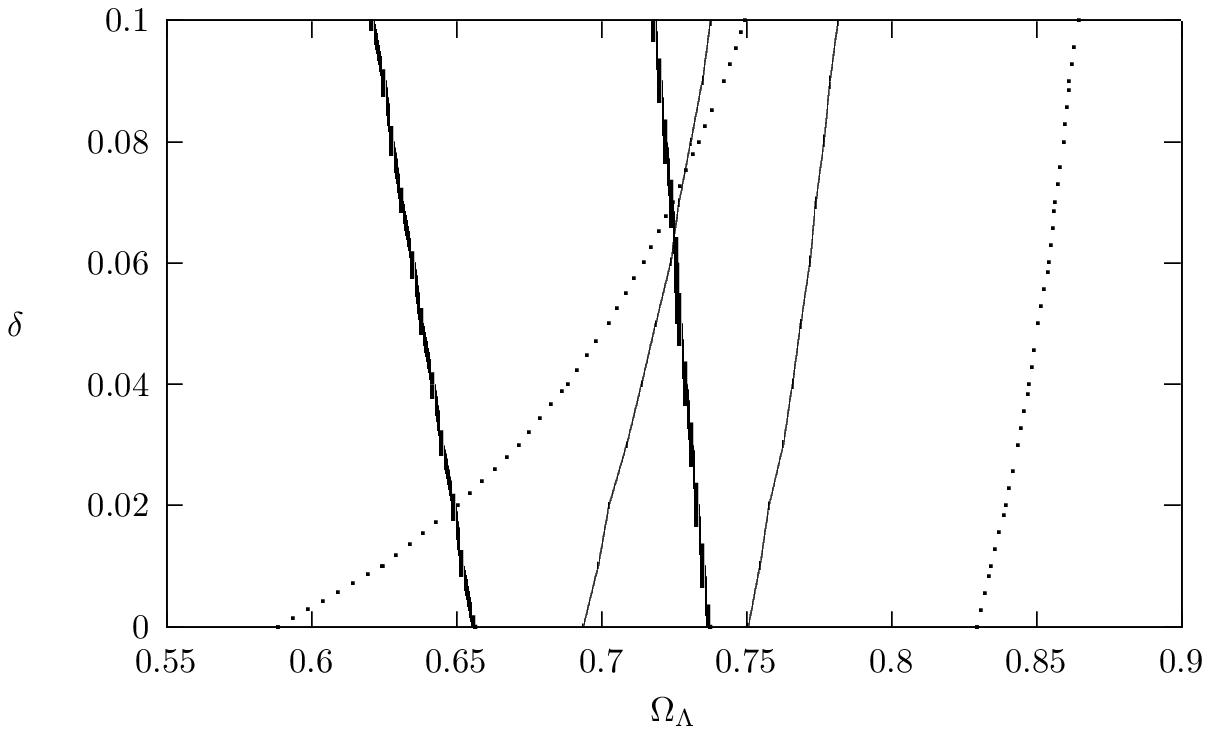}\\
\includegraphics[scale=1]{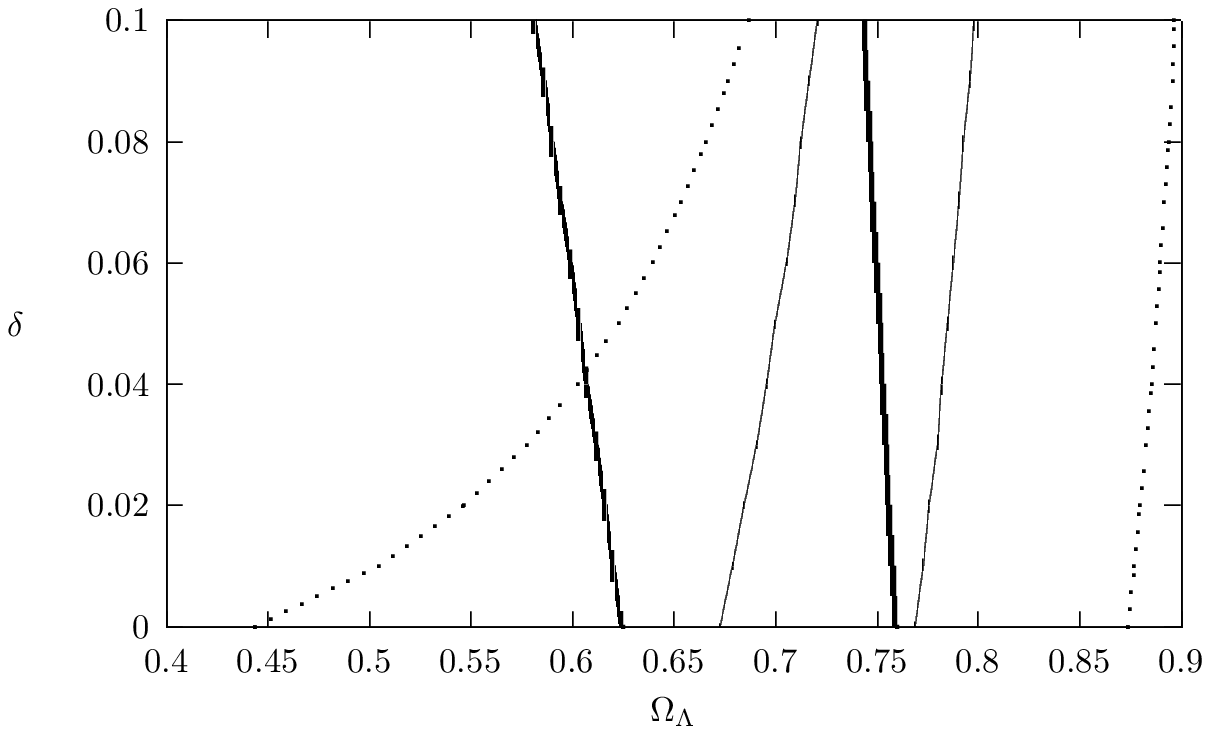}\\
\caption{The 68.3\% (upper panel) and 95.4\% (bottom panel) confidence level contours in the $\Omega_\Lambda$ - $\delta$ parameter space from the analysis of SNe (solid line), $H(z)$ (bold solid line) and BAO (dotted line) data, respectively.}
\end{figure}

\begin{figure}
  \includegraphics[scale=1]{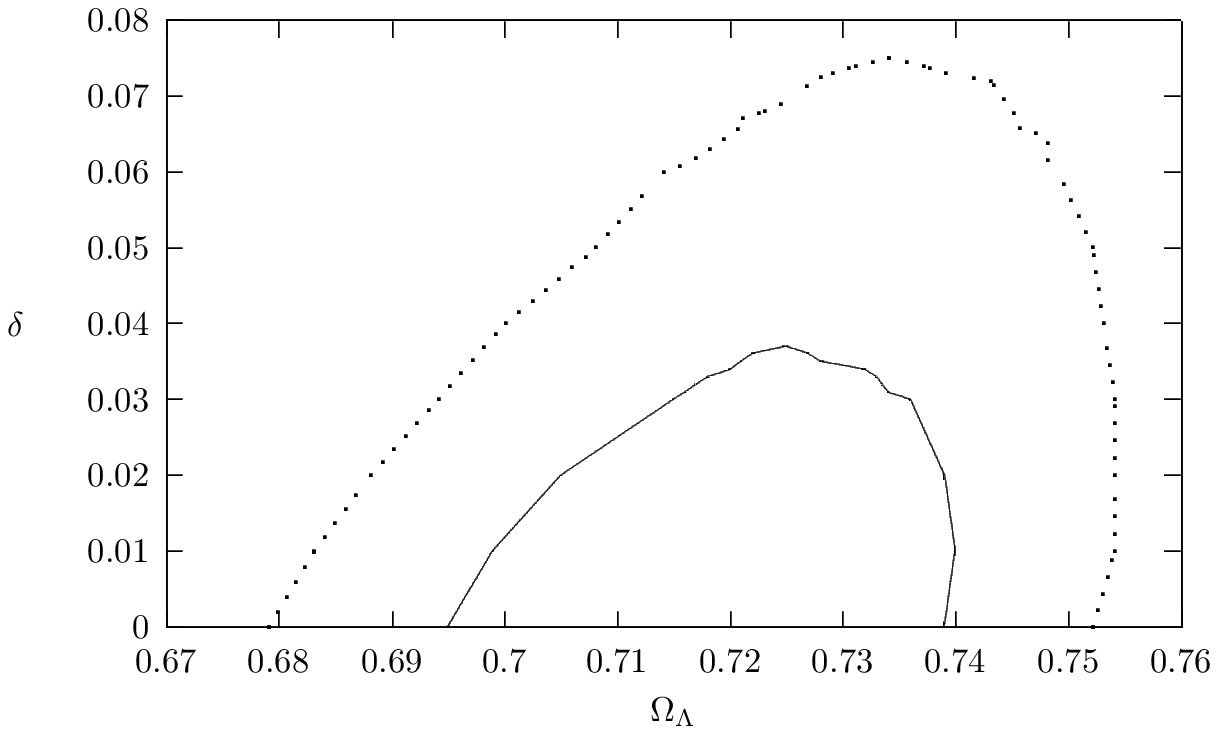}\\
  \caption{The 68.3\% (solid) and 95.4\% (dotted) confidence level contours in the $\Omega_\Lambda$ - $\delta$ parameter
space from the SNe+$H(z)$+BAO+matter density perturbation data analysis. The best-fit parameters for the observational data are $\Omega_{\Lambda}=0.712$, $\delta=0$ with
$\chi^{2}=562.39$.}
\end{figure}

\section{Other dark energy models on the brane}

Let us now consider the following model with a quite simple EoS,
\be\label{LR}
g(\rho_{D})=\alpha^{2}\rho_{D0}\left(\frac{\rho_{D}}{\rho_{D0}}\right)^{\beta},
\ee
where $\alpha$ and $\beta$ are dimensionless constants. If $\beta=1$, this reduces to an ordinary phantom energy model with constant EoS parameter $w=-1-\alpha^{2}$. From Eqs.~(\ref{brtrho}) and (\ref{arho})  one can see that, for various values of $\beta$, the model (\ref{LR}) describes three types of future universe evolution:

(a) Little Rip, for $\beta\leq 0$,

(b) Big Rip, for $0<\beta\leq 1$, and

(c) a type III singularity, for $\beta>1$.

Note that in the FRW cosmology the Little Rip occurs for $\beta\leq 1/2$. Simple calculations allow us to obtain the function $F(z)$:
\be \label{RHOZLR}
F(z)=\left\{\begin{array}{cc}\left[1-3\alpha^{2}(1-\beta)\ln(1+z)\right]^{\frac{1}{1-\beta}},\quad \beta\neq 1,\\
(1+z)^{-3\alpha^{2}},\quad \beta=1.
\end{array}\right.
\ee
We consider the case $\beta=0$ (Little Rip) and $\beta=1$.

The analysis of observational data for the Little Rip model leads to the same conclusions as in the case of the $\Lambda$CDM model: when the brane tension decreases, the common area of the confidence level contours for SNe, H(z), and BAO data in the $\alpha^{2}$ - $\Omega_{D0}$ parameter space decreases too, that is, the agreement with observational data becomes worse. In Fig.~3 the $1\sigma$ confidence level contours from the data set analysis are shown. The results for the joint observational data analysis are depicted in Fig.~4. One can see that, for large values of $\delta$, the description of the observational data is better for larger $\alpha^{2}$.

\begin{figure}
  \includegraphics[scale=0.8]{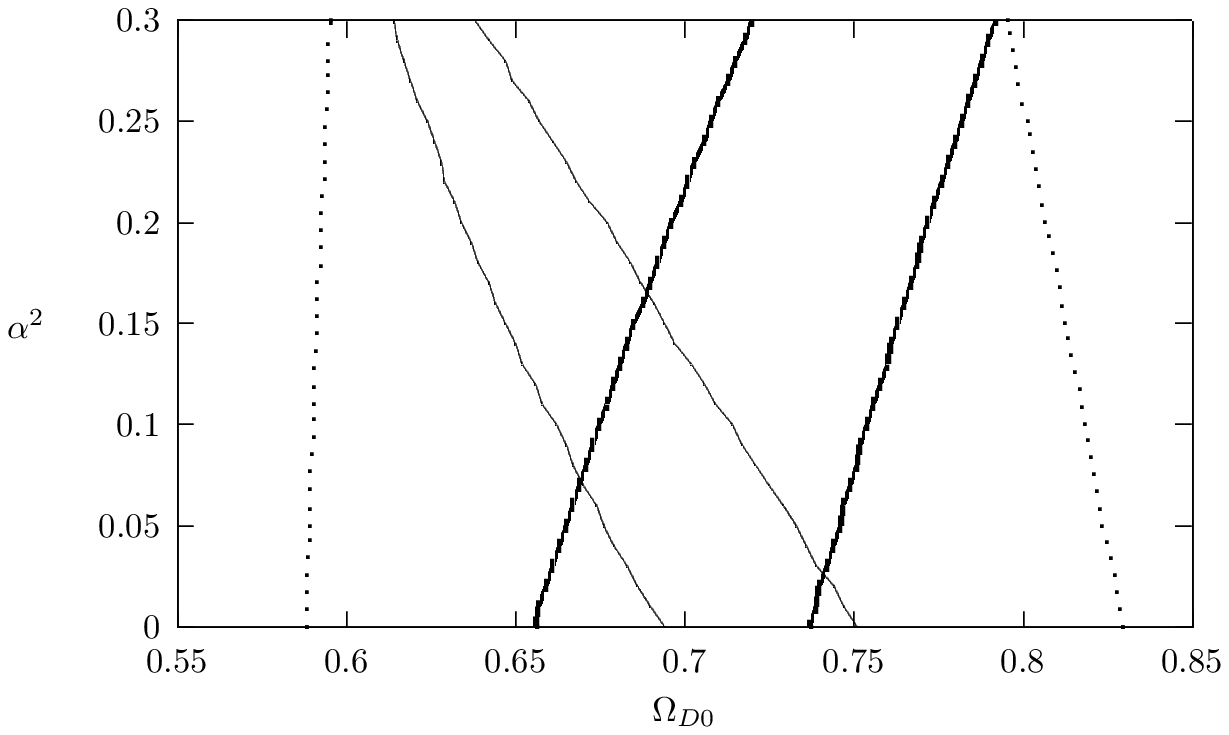}\\
\begin{center}{A}
\end{center}
  \includegraphics[scale=0.8]{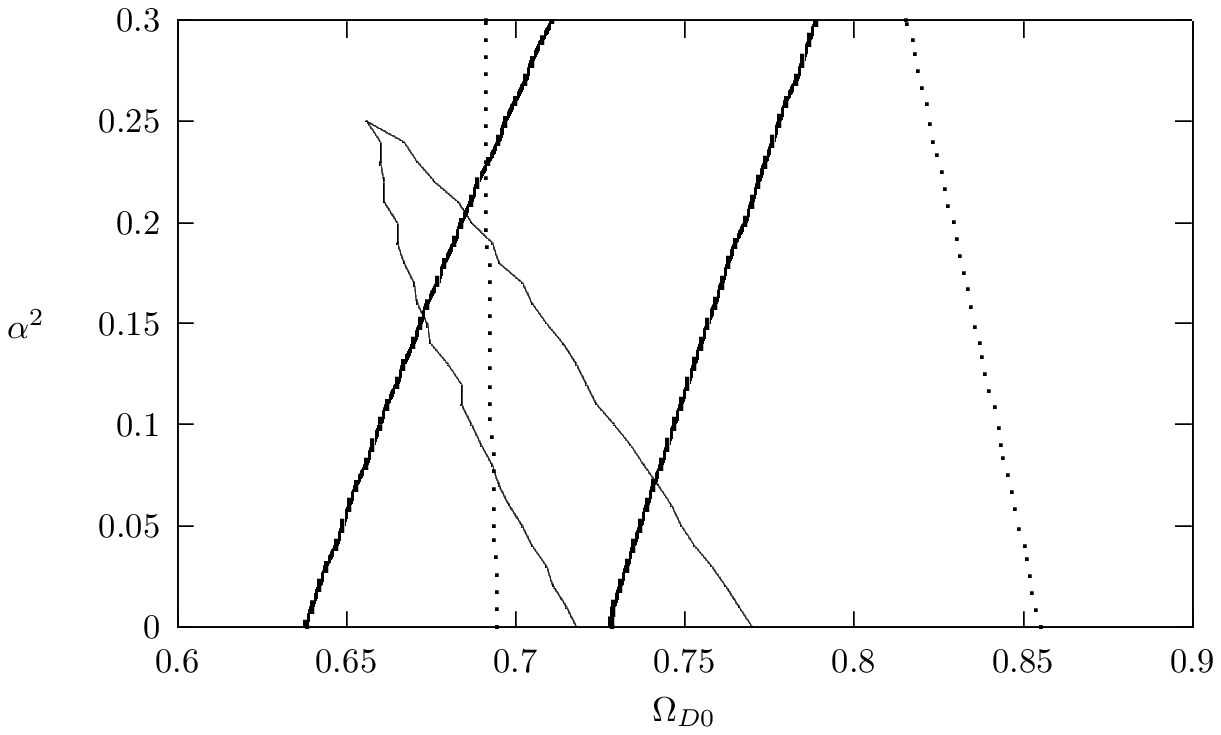}\\
  \begin{center}
  {B}
  \end{center}
  \includegraphics[scale=0.8]{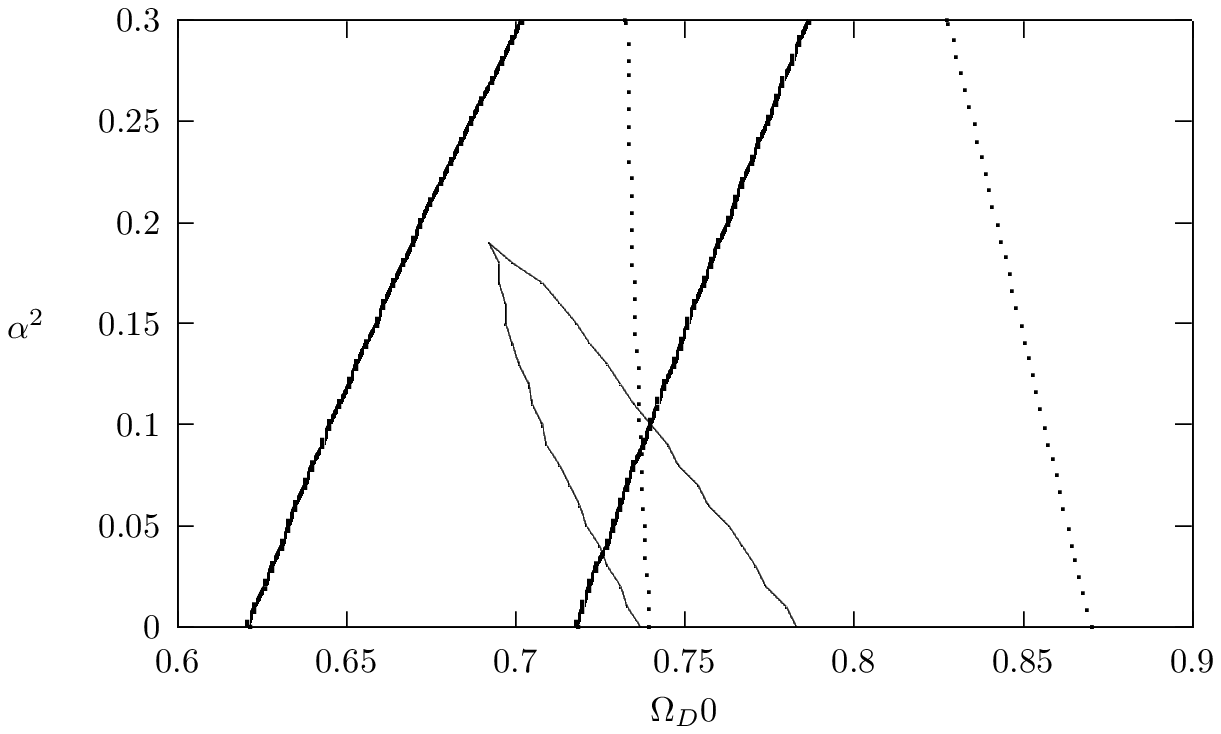}\\
  \begin{center}
  {C}
  \end{center}
  \caption{The 68.3\% confidence level contours in the $\Omega_\Lambda$ - $\alpha^{2}$ parameter space from the
  analysis of SNe (solid), $H(z)$ (bold solid), and BAO (dotted) data for the FRW cosmology (A), $\delta=0.05$ (B), and
  $\delta=0.10$ (C), in the case of the Little Rip model with $\beta=0$. We see that, for each of the data sets, the best-fit parameters correspond to the $\Lambda$CDM model ($\alpha^{2}=0$).}
\end{figure}

\begin{figure}
\includegraphics[scale=0.8]{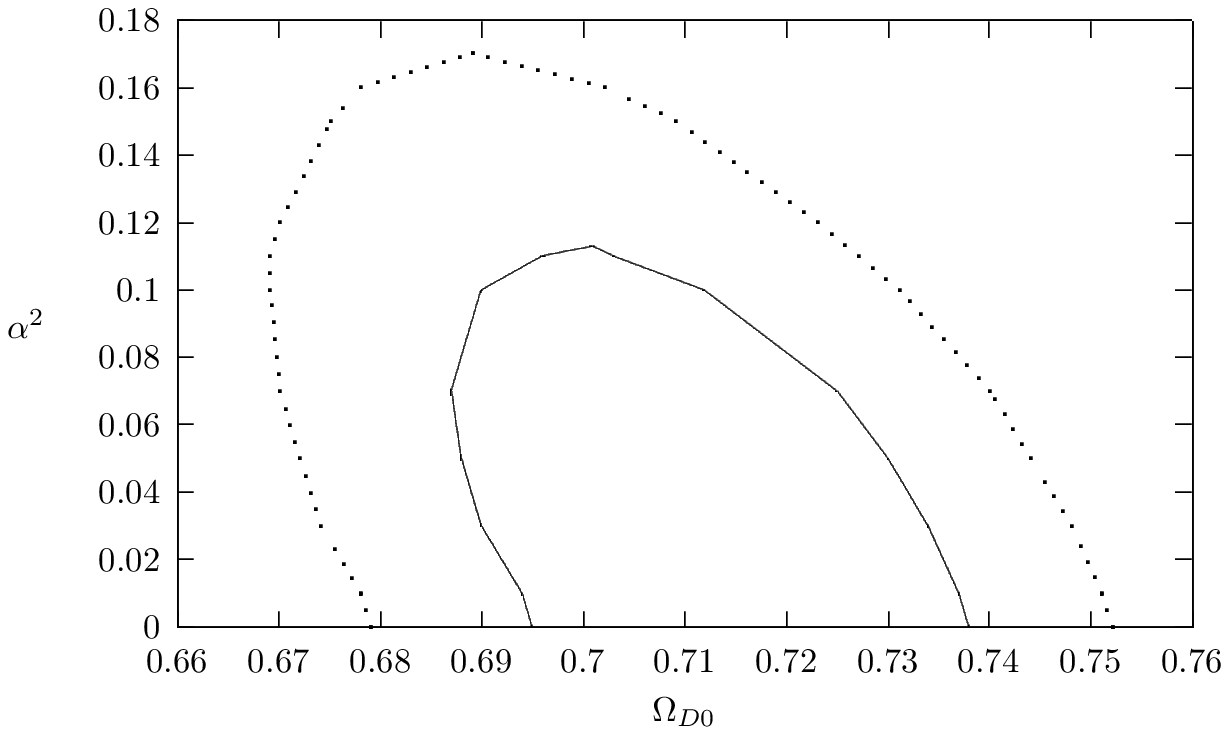}\\
\begin{center} {A} \end{center}
\includegraphics[scale=0.8]{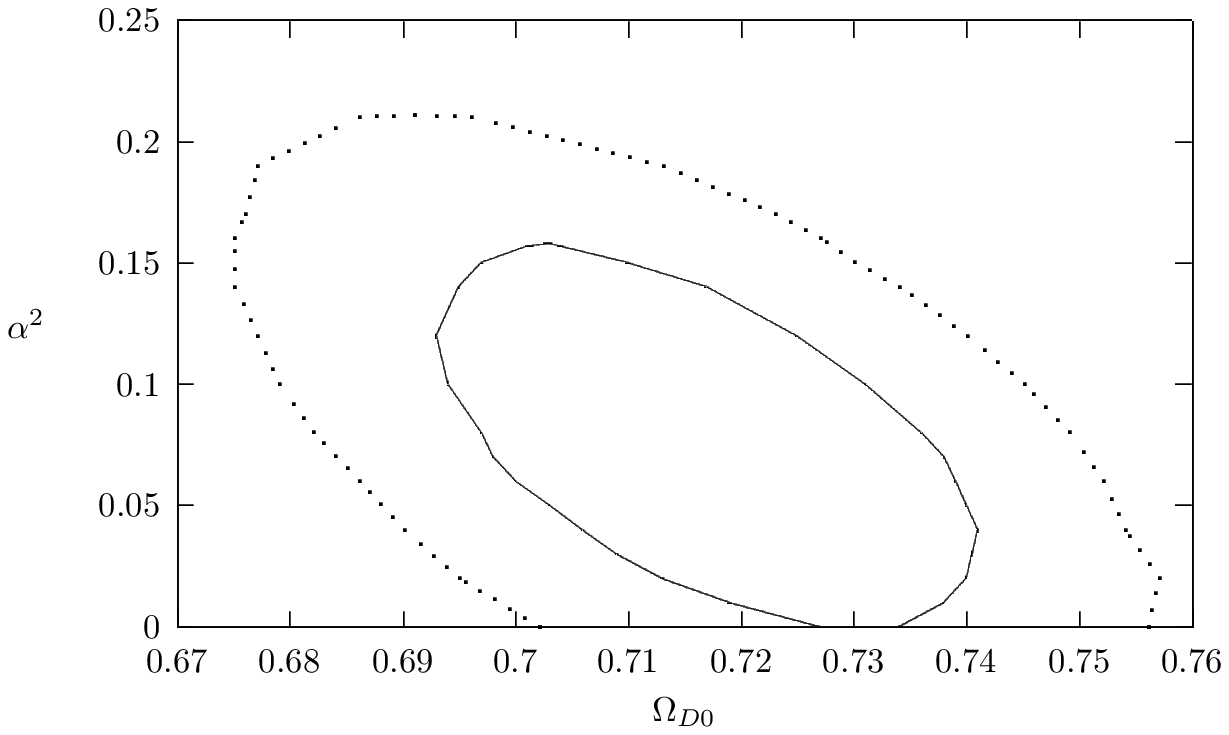}\\
\begin{center} {B} \end{center}
\includegraphics[scale=0.8]{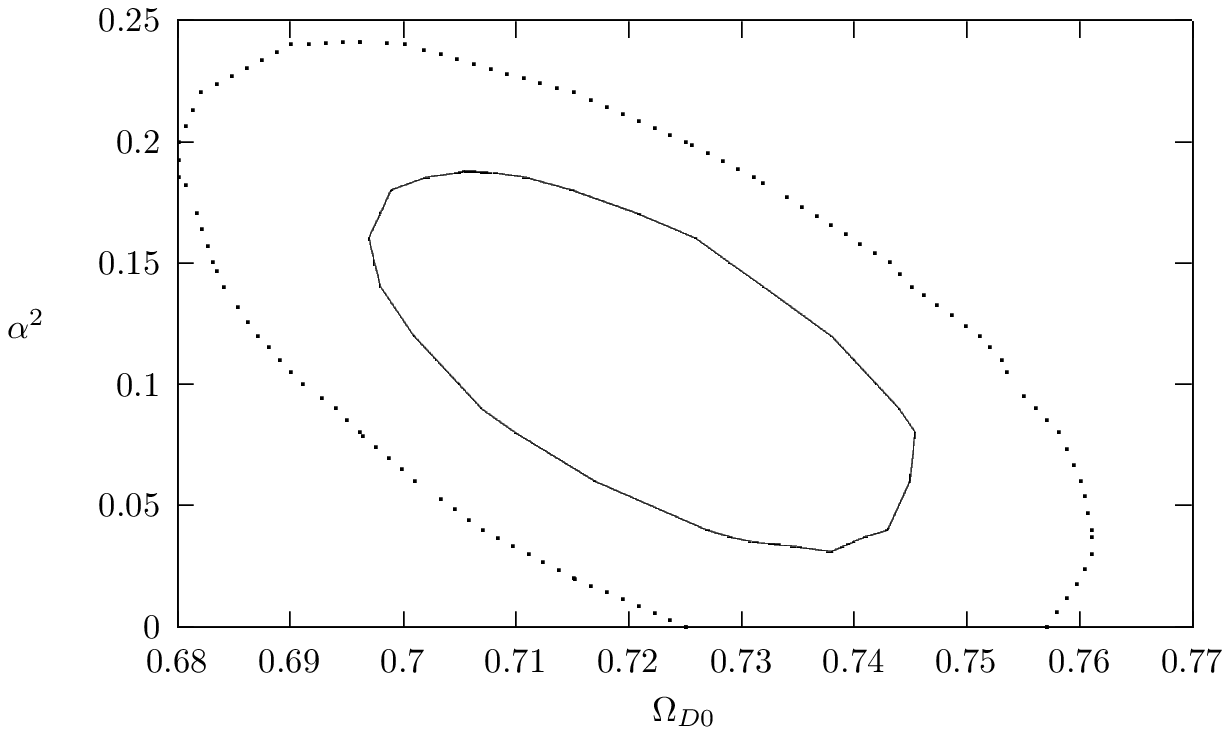}\\
\begin{center} {C} \end{center}
\caption{The 68.3\% (solid) and 95.4\% (dotted) confidence level contours in the $\Omega_\Lambda$ - $\alpha^{2}$
parameter space from the analysis of SNe+$H(z)$+BAO+matter density perturbation data for the FRW cosmology (A), $\delta=0.05$ (B), and
$\delta=0.10$ (C), in the case of the Little Rip model with $\beta=0$. The best-fit parameters are ($\alpha^{2}=0.03$, $ \Omega_{D0}=0.713$) with $\chi^{2}=562.20$ for the FRW cosmology, ($\alpha^{2}=0.08$, $\Omega_{D0}=0.714$) with $\chi^{2}=563.73$
for $\delta=0.05$, and ($\alpha^{2}=0.10$, $\Omega_{D0}=0.722$) with $\chi^{2}=566.78$ for $\delta=0.10$.}
\end{figure}

\begin{figure}
\includegraphics[scale=0.8]{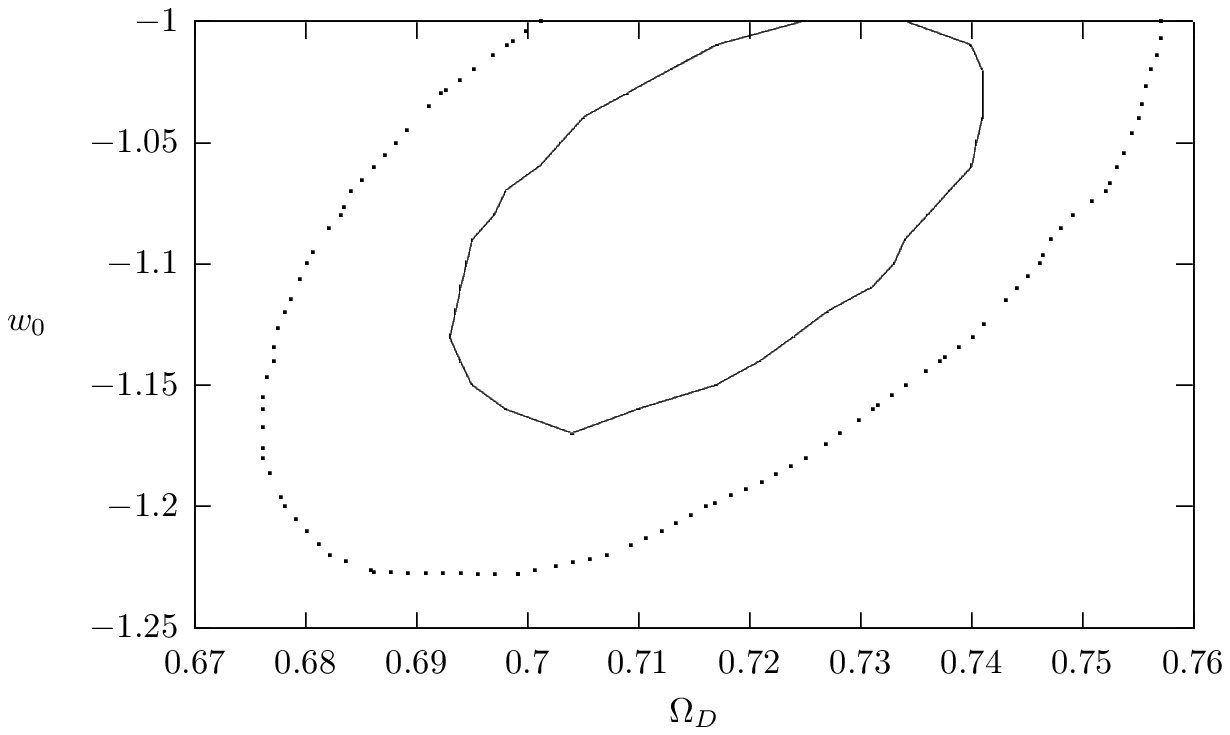}\\
\caption{The 68.3\% (solid) and 95.4\% (dotted) confidence level contours in the $\Omega_\Lambda$ - $w_{0}$  parameter space from the analysis of SNe+$H(z)$+BAO+matter density perturbation data for $\delta=0.05$ in the case of the simplest phantom model. The best-fit parameters are ($w_{0}=-1.08$, $\Omega_{D0}=0.718$) with $\chi^{2}=563.84$.}
\end{figure}

Similar results can be derived for the simplest phantom model with constant EoS parameter $w_{0}$. The best-fit parameters for the
FRW cosmology are ($w_{0}=-1.01$, $\Omega_{D0}=0.713$) with $\chi^{2}=562.16$ (recall, for comparison, that in the $\Lambda$CDM model we have the slightly larger value of $\chi^{2}_{min}=562.38$). In Fig.~5 the results of the joint analysis are depicted. For $\delta>0$ the observational data analysis speaks evidently in favor of $w_{0}<-1$.

\newpage

\section{State-finder parameters in teleparallel theories }

In this Section we will compute the so-called {\it state-finder} parameters in universes described by teleparallel models which may be considered as an example of Loop Quantum Cosmology. The first of these parameters is the effective $\omega$ parameter and the second the
deceleration parameter. Both are well-known in the literature.
The other two parameters, introduced due to increase in the accuracy of cosmological
data, were given in \cite{ssaa03} (see also \cite{bmno12}), with the purpose to advance 
beyond the effective $\omega$ and deceleration parameters. The four parameters are defined as follows:
\begin{enumerate}
 \item
The effective $\omega$ parameter is
\begin{eqnarray}\label{h1}
\omega_{eff}=-1-\frac{2\dot{H}}{3H^2};\end{eqnarray}

\item
the deceleration parameter
\begin{eqnarray}\label{h2}
q_{dec}=-\frac{1}{aH^2}\ddot{a}=-\left(\frac{\dot{H}}{H^2}+1 \right);\end{eqnarray}

\item
the jerk parameter
\begin{eqnarray}\label{h3}j=\frac{1}{aH^3}\dddot{a}=\frac{\ddot{H}}{H^3}+
3\frac{\dot{H}}{H^2}+1;
\end{eqnarray}
\item
and the snark parameter
\begin{eqnarray}\label{h4}s=\frac{j-1}{3(q_{dec}-1/2)}.
\end{eqnarray}
\end{enumerate}

We consider again a universe filled by a perfect fluid with EoS
$p=g(\rho)\equiv-\rho-f(\rho)$. Teleparallel theories in
flat FRW cosmology are defined via a Lagrangian of the form
${\mathcal L}_{T}=VF(T)-V\rho$ (see \cite{hhkn76,hs79,bho12} for a
review of the topic), where $V$ is the volume of the spatial part
and $T=-6H^2$ is the so-called {\it scalar torsion}
\cite{bf09,bmno12}.

From this Lagrangian we can see that the conjugate momentum of $V$
is  given by $p_V=\frac{\partial {\mathcal
L}}{\partial\dot{V}}= -4HF'(T)$, and thus the Hamiltonian is
\begin{eqnarray}\label{h5}
{\mathcal H}=
\dot{V}p_V- {\mathcal L}=  \left[ 2TF'(T)-F(T)  +\rho \right] V.\end{eqnarray}
It is well-known that  in general relativity  the Hamiltonian is constrained to be zero. This constrain leads to the modified Friedmann equation
\begin{eqnarray}\label{h6}
\rho=-2 F'(T)T+F(T)\equiv G(T),
\end{eqnarray}
which is a curve on the plane $(H,\rho)$.

Conversely,
given a curve of the form $\rho=G(T)$
for some function $G$, it could be obtained from the modified Friedmann
equation,  by choosing \cite{bho12}
\begin{eqnarray}\label{h7}
F(T)=-\frac{\sqrt{-T}}{2}\int \frac{G(T)}{T\sqrt{-T}}dT.
\end{eqnarray}
The modified Raychaudhuri equation is obtained from the modified Friedmann equation by
taking its derivative with respect  to  time and using the conservation equation $\dot{\rho} = 3Hf(\rho) $, from where $\dot{H}=-\frac{f(\rho)}{4G'(T)}$.
Then, the dynamics of the universe is given by the modified Raychaudhuri equation and the conservation equation, i.e. by the system
\begin{eqnarray}\label{h8}
\left\{\begin{array}{ccc}
\dot{H} &{=}& -\frac{f(\rho)}{4G'(T)},\\
\dot{\rho} &{=}& 3Hf(\rho),
\end{array}
\right. \end{eqnarray}
provided the universe moves along the curve $\rho=G(T)$.

To compute the state-finder parameters, one has to use
 the modified Friedmann, Raychauduri and conservation equations, to get
\begin{eqnarray}\label{h9}
\omega_{eff}(T)=-1-\frac{f(G(T))}{T G'(T)},\quad
q_{dec}(T)=-\left(\frac{3f(G(T))}{2TG'(T)}+1 \right),\nonumber\\
j(T)=\frac{9f(G(T))}{2T(G'(T))^3}\left[(f'(G(T))+1)(G'(T))^2-G''(T)f(G(T))\right]+1,
\nonumber\\
s(T)=-\frac{f(G(T))}{(G'(T))^2}\frac{[(f'(G(T))+1)(G'(T))^2-G''(T)
f(G(T))]}{f(G(T))+TG'(T)}.\end{eqnarray}
These formulas mean that the parameters are functions of $H$.
Since $H_0$ (the current value of the Hubble parameter) is
well-known, one can test all the $F(T)$ models
with current observations without actually solving them. In particular, as
we will see, one can test loop quantum cosmology or brane
cosmology in the Randall-Sundrum scenario, i.e., brane cosmology described in Sect.~II. In fact, $H_0$ could be calculated from measurements of
the luminosity distance $D_L(z)$, via the well-known formula
(\cite{ssaa03})
\begin{eqnarray}\label{h10}
H(z)=\left[\frac{d}{dz}\left(\frac{D_L(z)}{1+z}\right)\right]^{-1},
\end{eqnarray}
being $z$ the redshift, or either as we have already explained in Sect.~III B.

A remark is important: the formulas (\ref{h9}) could be expressed in function of $\rho$ via the relation $T=G^{-1}(\rho)$. This will
be relevant when we deal with
loop quantum cosmology and brane cosmology. These expressions read
\begin{eqnarray}\label{h9a}
\omega_{eff}(\rho)=-1-\frac{f(\rho)(G^{-1})'(\rho)}{G^{-1}(\rho)},\quad
q_{dec}(\rho)=-\left(\frac{3f(\rho)(G^{-1})'(\rho)}{2G^{-1}(\rho)}+1 \right),\nonumber\\
j(\rho)=\frac{9f(\rho)}{2G^{-1}(\rho)}\left[(f'(\rho)+1)(G^{-1})'(\rho)+
f(\rho)(G^{-1})''(\rho)\right]+1,
\nonumber\\
s(\rho)=-\frac{f(\rho)}{f(\rho)(G^{-1})'(\rho)+G^{-1}(\rho)}
[(f'(\rho)+1)(G^{-1})'(\rho)+f(\rho)(G^{-1})''(\rho)].\end{eqnarray}
We start  calculating the parameters for the simplest but one of
the most interesting EoS, when the dependence between pressure and
energy density is linear, i.e., for
 $$p=\omega\rho\Longleftrightarrow f(\rho)=-(1+\omega)\rho.$$

In that case, one has
\begin{eqnarray}\label{h11}
\omega_{eff}(T)=-1+(1+\omega)\frac{G}{TG'}, \quad
q_{dec}(T)=-\left(-\frac{3}{2}(1+\omega)\frac{G}{TG'}+1\right),\nonumber\\
j(T)=-\frac{9}{2}(1+\omega)\frac{G}{T(G')^3}\left(-\omega (G')^2 +(1+\omega)GG''  \right)+1,\nonumber\\
s(T)=\frac{(1+\omega)G[-\omega G' +(1+\omega)GG'']}
{(G')^2[-(1+\omega)G+TG']}.
\end{eqnarray}
As an application, we study FRW cosmology with a small cosmological constant $\Lambda$.
In this case $G(T)=-\frac{T}{2}-\Lambda$ and, from (\ref{h11}), a simple calculation yields
\begin{eqnarray}\label{h12}
\omega_{eff}(T)=\omega+2(1+\omega)\frac{\Lambda}{T},
\quad
q_{dec}(T)=-\left(-\frac{3}{2}(1+\omega)\left[1+\frac{2\Lambda}{T}\right]+1\right),\nonumber\\
j(T)=\frac{9}{2}(1+\omega)\omega\left[1+\frac{2\Lambda}{T}\right]+1,
\quad
s(T)={\omega}\left({1-\frac{1}{(1+\omega)\left[1+\frac{2\Lambda}{T}\right]}}\right)^{-1}.
\end{eqnarray}
We should remark that, when $\omega>-1$, at late times $\rho\rightarrow 0$ and thus  $T\rightarrow -2\Lambda$.  Obviously one has
$\omega_{eff}(T)\rightarrow -1$, $q_{dec}(T)\rightarrow -1$, $j(T)\rightarrow 1$ and $s(T)\rightarrow 0$. On the other hand,
 when $\omega<-1$ at late times one has a Big Rip singularity.

Recently, another interesting    model in FRW cosmology has  been introduced in
\cite{ha12} in order to deal with non-singular universes. One
considers once again the curve $\rho=G(T)=-\frac{T}{2}-\Lambda$,
but with a non-linear EoS
\begin{equation}\label{h20}
p(\rho)=-\frac{\rho^2}{\rho_i}  \Leftrightarrow f(\rho)=-\rho\left(1-\frac{\rho}{\rho_i}\right),
\end{equation}
where $\rho_i$ is a constant satisfying $\Lambda\ll \rho_i$.

This model has two de Sitter solutions $H_f=\sqrt{\frac{\Lambda}{3}}$ and $H_i=\sqrt{\frac{\Lambda+\rho_i}{3}}$, and
shows a universe evolving from an early  inflationary phase (the de Sitter phase $H_i$) to late time accelerated expansion
(de Sitter phase $H_f$), going trough a matter dominated phase which allows the formation of structures. It could be
also viewed as a universe with a huge cosmological constant $\rho_i$ at early times, which
evolves, at late times, towards a small cosmological constant $\Lambda$ responsible of the current cosmological acceleration.

In this case, for the current (small) value of $T$, using (\ref{h9a}), (\ref{h20}) and $G^{-1}(\rho)=-2(\rho+\Lambda)$, one gets
\begin{eqnarray}\label{h21}
\omega_{eff}(\rho)=-1-\frac{\rho(\rho_i-\rho)}{\rho_i(\rho+\Lambda)}\quad
q_{dec}(\rho)=-1-\frac{3\rho(\rho_i-\rho)}{2\rho_i(\rho+\Lambda)},\nonumber\\
j(\rho)=1-\frac{9\rho^2(\rho_i-\rho)}{\rho_i^2(\rho+\Lambda)},\quad
s(\rho)=-2\frac{\rho^2(\rho_i-\rho)}{\rho_i(\rho^2+\rho_i\Lambda)}.
\end{eqnarray}

\subsection{Loop Quantum Cosmology and Brane Cosmology with a small cosmological constant $\Lambda$}

In loop quantum cosmology, which reminds of brane cosmology, the effective Friedmann equation depicts
 the following ellipse (see, for example, \cite{bho12})
\begin{eqnarray}\label{h13}
H^2=\frac{\rho+\Lambda}{3}\left(1-\frac{\rho+\Lambda}{\rho_c}  \right),
\end{eqnarray}
on the plane $(H,\rho)$, where $\rho_c$ is the so-called {\it critical density}, that  satisfies $\Lambda\ll \rho_c$.
This curve can be written in two pieces, $\rho_{m}=G_-(T)$ and $\rho_{m}=G_+(T)$, where
\begin{eqnarray}\label{h14}
G_{\pm}(T)=-\Lambda+\frac{\rho_c}{2}\left(1\pm\sqrt{1+\frac{2T}{\rho_c}} \right).\end{eqnarray}

Since nowadays $H_0$ and $\rho_0$ have small values, we need to  choose $G(T)\equiv G_-(T)$ and then, using the formula (\ref{h7}), we get
\begin{eqnarray}\label{h15}
 F(T)=-\sqrt{-\frac{T\rho_c}{2}}\arcsin\left(\sqrt{-\frac{2T}{\rho_c}}\right)+
\frac{\rho_c}{2}\left(1-\sqrt{1+\frac{2T}{\rho_c}}
\right)-\Lambda,
\end{eqnarray}
what shows that the effective formulation of LQC is a teleparallel theory.

To compare with FRW cosmology, since nowadays $T$ is small as compared with $\rho_c$, we can expand $G(T)$ up to second order in $T$, to get
\begin{eqnarray}\label{h16}
G(T)=-\Lambda-\frac{T}{2}+\frac{T^2}{4\rho_c},
\end{eqnarray}
and inserting this expression into (\ref{h9}), one obtains the first order correction to the FRW cosmology.
Moreover, in order to obtain exact formulas, one has to use Eq.~(\ref{h9a}), because in that case $G^{-1}(\rho)$ has a very simple, quadratic
expression, namely
$$
G^{-1}(\rho)=-2({\rho+\Lambda})\left(1-\frac{\rho+\Lambda}{\rho_c}\right).
$$
A straightforward calculation yields
\begin{eqnarray}\label{h17}
\omega_{eff}(\rho)=-1+\frac{f(\rho)(\rho_c-2(\rho+\Lambda))}{(\rho+\Lambda)(\rho_c-(\rho+\Lambda))}, \quad
q_{dec}(\rho)=-1+\frac{3}{2}\frac{f(\rho)(\rho_c-2(\rho+\Lambda))}{(\rho+\Lambda)(\rho_c-(\rho+\Lambda))},\nonumber\\
j(\rho)=-\frac{9}{2}\frac{f(\rho)[-(f'(\rho)+1)(\rho_c-2(\rho+\Lambda))+2f(\rho)]}{(\rho+\Lambda)(\rho_c-(\rho+\Lambda))}+1
,\nonumber\\
s(\rho)=\frac{f(\rho)[-(f'(\rho)+1)(\rho_c-2(\rho+\Lambda))+2f(\rho)]}{(\rho+\Lambda)(\rho_c-(\rho+\Lambda))+f(\rho)(\rho_c-2(\rho+\Lambda))}
.
\end{eqnarray}

On the other hand,
as we already saw in Sect.~II, in brane cosmology in the Randall-Sundrum scenario  the modified Friedmann equation depicts
 the following  hyperbola
\begin{eqnarray}\label{h18}
H^2=\frac{\rho+\Lambda}{3}\left(1+\frac{\rho+\Lambda}{2\lambda}  \right),
\end{eqnarray}
on the plane $(H,\rho)$.

Finally, by comparing this equation with (52) and making
the change $\rho_c\rightarrow -2\lambda$, it follows that one can view loop quantum
cosmology as brane cosmology with a negative brane tension and a time-like
extra dimension. Therefore, in order to obtain the corresponding state-finder
parameters formulas  in brane cosmology, we just need to do the
replacement $\rho_c\rightarrow -2\lambda$. Moreover, with this replacement we can apply the general formulas (\ref{h17}) to any EoS, in particular to the model studied in Sect.~IV.

A very important remark is here in order. As a result of the above replacement it follows that the dynamics resulting for both theories, LQC and BC,
are very different, because their corresponding Freedmann equations
 depict
two completely different curves. In particular, Rip singularities, as we have seen in Sect.~V, are allowed in BC because the hyperbola is an unbounded curve. But, since in LQC
the Friedmann equation depicts a bounded curve (an ellipse), Rip singularities cannot appear in this case.
For example, for the EoS $p(\rho)=\omega\rho$ the universe is
non-singular (see \cite{ha12} for a detailed explanation). In
fact, for $\omega>-1$ (resp. $\omega<-1$)
 it moves in  anti-clockwise (resp. clockwise) way
from the anti de Sitter solution
$H=-\sqrt{\frac{\Lambda}{3}}\sqrt{1-\frac{\Lambda}{\rho_c}}$
(resp. de Sitter solution
$H=\sqrt{\frac{\Lambda}{3}}\sqrt{1-\frac{\Lambda}{\rho_c}}$) towards
the de Sitter one
$H=\sqrt{\frac{\Lambda}{3}}\sqrt{1-\frac{\Lambda}{\rho_c}}$ (resp.
 anti de Sitter one
$H=-\sqrt{\frac{\Lambda}{3}}\sqrt{1-\frac{\Lambda}{\rho_c}}$).

\begin{figure}
  \includegraphics[scale=1]{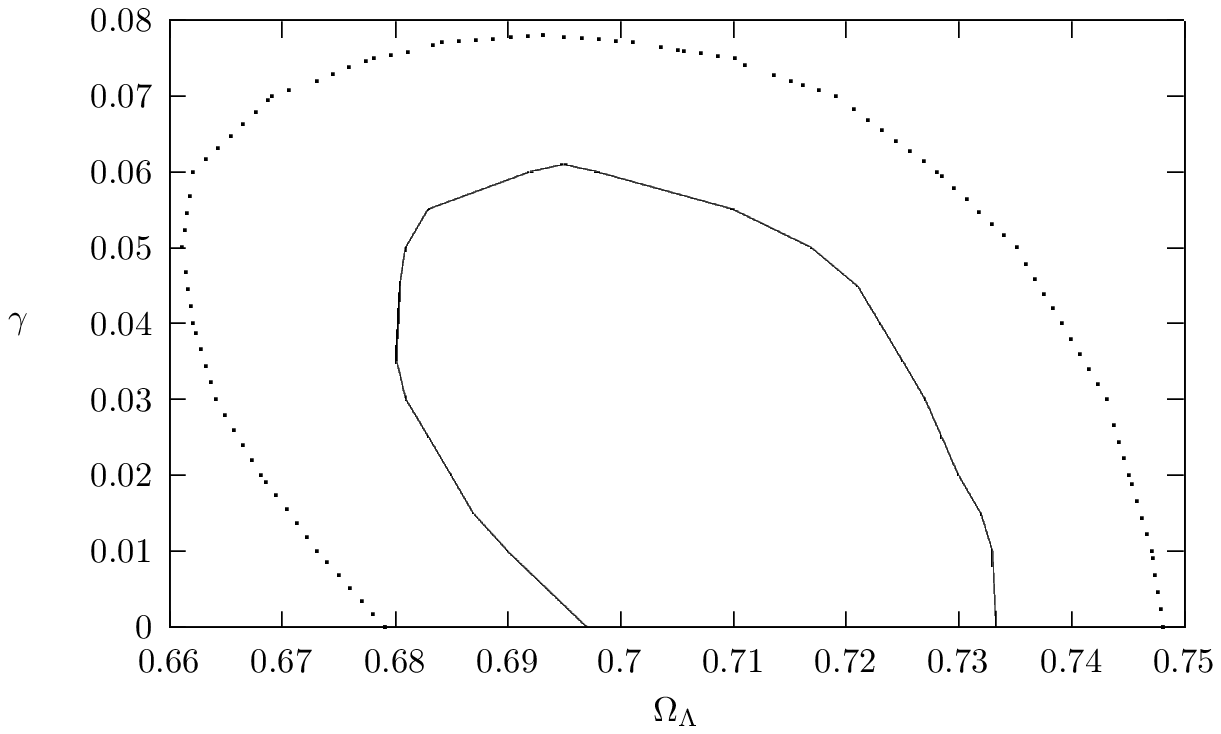}\\
  \caption{The 68.3\% (solid) and 95.4\% (dotted) confidence level contours in the $\Omega_\Lambda$ - $\gamma$ parameter
space from the SNe+$H(z)$+BAO data analysis in the case when $\rho_D=\Lambda$. The best-fit parameters for the observational data are $\Omega_{\Lambda}=0.712$, $\rho/\rho_{c}=0.03$ with $\chi^{2}=560.58$ (to be compared with $\chi^{2}_{min}=561.31$ for the $\Lambda$CDM model).}
\end{figure}

In the early universe the parameter $\rho_c$ is very large as compared with
$\rho_0$ ($\rho_c\gg \rho_0$). In principle, $\rho_c$ is fixed and has the value
$\rho_c=3/(\gamma_{BI}^2\lambda_{LQC}^2)$, being $\gamma_{BI}=0.2375$
the Barbero-Immirzi parameter and $\lambda_{LQC}^2=\sqrt{3}\gamma_{BI}/4$
(see, for instance, \cite{bho12}). But one could also assume that $\rho_c$ can vary
with time  or, in other words, we can consider LQC with the parameter $\rho_{c}$ in some cosmological epoch. For illustration we can estimate the possible values of $\rho/\rho_{c}$ at the present epoch (at redshifts $0<z<1.75$). As the LQC equations formally coincide with the cosmological equations on the brane in the RS scenario, one can confront the simplest loop quantum cosmology model ($\rho_D=\Lambda$) with observational data, in the same manner as it was done in Sect.~IV. The analysis shows (see Fig.~6) that the best fit for the SNe+$H(z)$+BAO data is achieved for $\gamma=(\rho_{0}+\Lambda)/\rho_{c}\approx 0.03$. Maybe this result can be considered as an argument in favor of loop quantum cosmology.

\section{Conclusions}

We have confronted in this paper several DE models on the brane with combined data coming from
different, totally independent cosmological surveys. The analysis here performed shows that the fit of
these observational data is actually better for the FRW cosmology in the framework of the chosen model for dark energy. Also, owing to the fact that the LQC equations formally coincide with the cosmological equations on the brane in the RS scenario, we could additionally confront the simplest (but very important) loop quantum cosmology model ($\rho_D=\Lambda$) with observational data, in the same manner as it was done with the brane models. The analysis we have carried out has shown (Fig.~6) that the best fit for SNe+$H(z)$+BAO data is achieved for $\gamma=(\rho_{0}+\Lambda)/\rho_{c}\approx 0.03$, what could be actually viewed as a good argument in favor of loop quantum cosmology.

Taking everything into account, the observational cosmological results do not exclude, in principle, that the real cosmology of the universe we live in could in fact differ from that of the standard FRW model. A window remains still open for discrepancy. The importance of joint analysis of the various, independent classes of observational
data sources has been clearly manifested in the discussion of the different tables and plots. Taking into account together SNe
apparent magnitude measurements, Hubble parameter evolution data, and BAO and matter density perturbation data, we are able to get a
quite rigid constraint on the allowed value of the brane tension, in the framework of the different brane models considered here.

\subsection*{Acknowledgements.}

The work of AVY has been supported by the ESF, project 4868 ``The cosmological constant as eigenvalue of Sturm-Liouville problem'', and the work of AVA has been supported by the ESF, project 4760 ``Dark energy landscape and vacuum polarization account'', both within the European Network ``New Trends and Applications of the Casimir Effect''. EE's research has been partly supported by MINECO (Spain), contract PR2011-0128, and it was partly carried out while on leave at the Department of Physics and Astronomy, Dartmouth College, 6127 Wilder Laboratory, Hanover, NH 03755, USA. JdH was supported by MINECO (Spain), project MTM2011-27739-C04-01, and AGAUR (Generalitat de Catalunya), contract 2009SGR-345.  EE and SDO have been supported in part by MICINN (Spain) project FIS2010-15640, by the CPAN Consolider Ingenio Project, and by AGAUR (Generalitat de Ca\-ta\-lu\-nya), contract 2009SGR-994.

\end{document}